\documentstyle[12pt,aasms4]{article}

\lefthead{Swaters, Sancisi \& van der Hulst}
\righthead{The \ion{H}{1} Halo of NGC 891}

\begin{document}
\title{The \ion{H}{1} Halo of NGC 891}
\author{R.A. Swaters, R. Sancisi, J.M. van der Hulst}
\affil{Kapteyn Astronomical Institute, University of Groningen, P.O. 
Box 800, 9700 AV, The Netherlands}

\begin{abstract}

Neutral hydrogen observations of the nearby, edge-on spiral galaxy
NGC~891 reveal the presence of an \ion{H}{1} halo extending up to at
least 5 kpc from the plane.  This halo gas appears to rotate 25 to 100
km/s more slowly than the gas in the plane.  If this velocity difference
is due to the gradient in the gravitational potential, then it may serve
to discriminate between disk and spheroidal mass models.  The classic
picture of a large outer flare in the \ion{H}{1} layer of NGC~891 may no
longer be valid. 

A correlation is seen between the distributions of \ion{H}{1},
H$\alpha$ and radio continuum emission, which supports, in accordance
with galactic fountain models, the picture of a substantial disk-halo
circulation related to the star formation activity in the disk of
NGC~891.

There is now also clear evidence for the presence of a rapidly rotating
($v_{rot} \simeq 230$ km/s) disk or ring of \ion{H}{1} in the central
part of NGC~891. 

\end{abstract}

\keywords{galaxies: individual (NGC 891) --- galaxies: ISM --- galaxies:
kinematics and dynamics}

\section{Introduction}

NGC~891 is a nearby Sbc galaxy, seen almost perfectly edge-on.  In
recent years it has been studied in detail at various wavelengths and
has become the key system for the study of the vertical distribution
of the interstellar medium in a disk galaxy.

Early studies with the Westerbork Synthesis Radio Telescope (WSRT)
showed the presence of a radio continuum halo (Allen, Baldwin, \&
Sancisi\ 1978) and of \ion{H}{1} emission at large distances from the
plane (Sancisi \& Allen 1979, hereafter SA), later confirmed by VLA
observations (Rupen 1991).  This emission was interpreted by SA as due
to a huge flare of the outer \ion{H}{1} disk of NGC 891. 

In the central region of NGC 891 the observations showed a deep
\ion{H}{1} depression, which was attributed, at least partly, to the
effect of \ion{H}{1} absorption.  A CO nuclear ring or disk was
discovered by Sofue, Nakai, \& Handa (1987) and has more recently been
confirmed and studied in detail by Garc\'{\i}a-Burillo et al.\ (1992,
hereafter GGCD) and Garc\'{\i}a-Burillo \& Gu\'elin (1995), but no such
central system was detected in \ion{H}{1} by SA or Rupen (1991). 

NGC 891 was re-observed with the WSRT for 12 times 12 hours between
March 1988 and January 1989.  This provided us with the most sensitive
\ion{H}{1} data for an edge-on galaxy to date.  Here we present the two
main findings of these observations: the \ion{H}{1} halo and the central
\ion{H}{1} ring or disk of NGC 891.  A more detailed description of the
observations and of the results will be presented elsewhere. 

\section{The \ion{H}{1} Halo}

The map of the total \ion{H}{1} column density superposed on the optical
image (Fig.\ 1) shows a thin, high-density \ion{H}{1} layer coinciding
with the dust lane in the galaxy plane and wings extending up to about 2
arcmin on both sides (1 arcmin $\simeq$ 2.8 kpc).  The total amount of
\ion{H}{1} above 30 arcsec is about $6\cdot 10^8\hbox{~M}_\odot$, or
$\sim 15$\% of the total \ion{H}{1} mass of NGC 891.  This \ion{H}{1} at
large $z$-distances from the plane, already detected (but with much
lower signal/noise ratio) by SA, had been interpreted as being part of a
flaring \ion{H}{1} layer at large radii.  This conclusion was based on
the kinematics of the \ion{H}{1}, shown here in the position-velocity
maps in the directions along (Fig.\ 2) and perpendicular (Fig.\ 3) to
the major axis of NGC 891.  The \ion{H}{1} emission at large
$z$-distances (Fig.\ 3) has, in fact, radial velocities close to the
systemic velocity of NGC 891 and should therefore, according to the
kinematical model adopted by SA, be located in the outer parts of the
system and hence be part of a large flare.  It is important to note that
in this standard model cylindrical rotation is assumed, i.e.  the gas at
a given radius in the galaxy has a rotational velocity independent of
its $z$-distance from the plane. 

With the new, higher sensitivity \ion{H}{1} observations and a more
sophisticated analysis that makes use of a three dimensional modeling,
we now have obtained a different, clearer picture of the vertical
structure of the \ion{H}{1} in NGC 891.  Models have been constructed
for a range of different inclination angles, outer flares and
line-of-sight warps.  The input values for the rotation curve and the
radial \ion{H}{1} density distribution have been derived directly from
the data.  The standard value of 10 km/s has been adopted for the
velocity dispersion.  The other parameters, disc scaleheight,
inclination and position angles, have been used to produce best fit
models for an \ion{H}{1} layer with flare, warp and various
inclinations.  The models have been inspected and analyzed in the same
way as the data cubes in order to make a full comparison with the 3-D
observational data set.  Some of the results are shown in Fig.\ 4
where each column shows representative channel maps for a given model.
The flare, warp and slow rotation models shown here are for
$i=90^\circ$.  The rightmost column shows the data for the north
(approaching) side of NGC 891.  The galaxy center is marked by the
cross on the right.  The top row shows the channel maps at the
systemic velocity, the bottom row shows channel maps at the extreme
velocity, close to the rotation velocity of the system.  These latter
channel maps, in standard circular rotation models, show the
\ion{H}{1} at the line of nodes of the orbits in the sky plane where
the line of sight is tangent to the orbits.

From Fig.\ 4, it is clear that the models obtained for a thin
(FWHM$=20''$) \ion{H}{1} layer viewed at different inclination angles
fail to reproduce the observations.  An inclination angle as low as
$80^\circ$ would be needed to reproduce the thick distributions in the
observed channel maps, but such a low inclination would be totally
inconsistent with the well-established lower limit of $\simeq
88\rlap{.}^\circ 6$ (cf.\ Rupen et al.\ 1987).  Furthermore, at such a
low inclination the V-shaped signature which is already noticeable at
$87^\circ$ is very prominent, whereas it is not seen in the data.  The
flare model does not reproduce the overall shape of the observed
channel maps, in particular not the remarkably thin structure of the
hydrogen in the extreme velocity channel.  This thin structure is
reproduced in the line-of-sight warp models, but these show a
different shape and a bifurcation that is not seen in the data at the
velocities closer to systemic. Undoubtedly, a warp is present in the
outer parts, as one can see from the data in Figs.\ 1 and 4: the gas
at $\pm 6'$ from the center is displaced to the west on the NE side of
the galaxy by about $10''$, and on the SW side it is displaced to the
east by about the same amount. It has the characteristic integral-sign
shape also seen in other edge-on galaxies, but it looks much less
pronounced. The test made with the present model is for a warp more
oriented in the line of sight, which, if present, would explain the
large symmetric $z$-extension. The models appear to rule out this
possibility. We have also modeled combinations of warp, inclination
and flare and none of these models was able to reproduce the data
satisfactorily.

All the models described above are based on the standard assumption that
the gas at large $z$ heights rotates with the same velocity as the gas
in the plane.  The column next to the observed channel maps in Fig.\ 4
shows a model in which this assumption has been dropped and the gas
layer is made up of two components, a thin disk (i.e.\ a Gaussian
$z$-distribution with a $\hbox{FWHM}=20''$) rotating according to the
derived rotation curve in the plane (Fig.\ 5), and a thick disk
(Gaussian with $\hbox{FWHM}=80''$) for which the flat part of the
rotation curve is 25 km/s lower (see lower curve in Fig.\ 5).  This
model gives a better reproduction of the observations: both the general
shape of the \ion{H}{1} distribution and especially the thin structure
in the channel map at the highest circular velocity (bottom row Fig.\ 4)
are reproduced.  A slight improvement is obtained if a larger value (20
to 30 km/s) is used for the velocity dispersion instead of the 10 km/s
adopted here.  In contrast to the flare model, this two-component model
has gas in the halo region (thick component) everywhere above the thin
disk.  But, because of its lower rotation velocity, the thick component
(or halo emission) does not show up in the extreme velocity channels as
these correspond to the higher rotation velocity of the thin disk. 

Such a model, with a thick component rotating more slowly than the thin
disk, is likely to be overly simplified.  It is more likely that there
is a gradual decrease of the density and of the rotation velocity with
increasing $z$ height.  In order to obtain an overall picture of such
variations we have derived a `tangent-velocity field' of the whole
galaxy, in and above the plane, with the assumption that there is
\ion{H}{1} everywhere, filling the orbits at the nodes (where the line
of sight velocity is maximal) for orbits of all radii.  This model is a
very simple configuration of stratified gas layers with no outer
flaring.  And, as usual for edge-on galaxies (cf.  SA), at each position
the rotation velocity is given by the velocity with the largest
deviation from systemic, after correction for instrumental broadening
and random motions.  Fig.\ 5 shows the rotation curve for the gas in the
plane (top curve) and the rotation curve above the plane (bottom curve). 
To derive the latter we have integrated over the strip $30''<z<60''$ and
also averaged on the two sides of the galaxy to increase the
signal/noise ratio.  In the inner region these rotation curves, mainly
because of the poor S/N ratio, are not well defined.  The
position-velocity diagram for the galactic plane, shown in Fig.\ 2,
indicates a very steep velocity rise near the center and a peak within 1
kpc radius, probably due to a central rotating disk or ring (see below). 
Farther out, between 2 and 5 kpc, there is a dip and a new rise to the
flat part ($v_{rot}=225$ km/s) of the curve.  The rotation curve (Fig.\
5) for the gas above the plane, derived under the assumption that there
is gas at the line of nodes, is indeed, in its flat part, about 25 km/s
lower.  In the inner regions, between 4 and 10 kpc, the difference is
much larger, up to 100 km/s.  The rotation velocity (flat part) seems to
decrease smoothly with increasing $z$-distance from the plane (cf.\
Fig.\ 3).  A rough estimate of the vertical gradient is 20 km/s per kpc. 

\section{Discussion}

The principal, new findings of these observations of NGC 891 are the
\ion{H}{1} halo and its slower rotation as compared to the \ion{H}{1} in
the plane.  New questions now arise.  What is the origin of this halo
gas? What causes its slower rotation? Is there still, to some extent, a
flare in the outer \ion{H}{1} layer?

There is little doubt that the origin of this halo gas and its kinetic
energy are related to the star formation activity in the disk of NGC 891
which is particularly intense, as indicated by the strong H-alpha (Rand,
Kulkarni, \& Hester 1990) and the radio continuum emission (Allen et
al.\ 1978; Dahlem, Dettmar, \& Hummel 1994).  These components are, like
the \ion{H}{1}, quite extended in the $z$-direction.  More recently,
also diffuse hot X-ray gas has been found (Bregman \& Pildis 1994). 
Further support of the view that a high star formation rate in the disk
lies at the origin of a disk-halo gas flow comes from \ion{H}{1}
observations of galaxies seen face-on.  In NGC 6946, where the star
formation activity is also high, widespread high velocity \ion{H}{1}
with vertical motions of up to 100 km/s has been detected by Kamphuis \&
Sancisi (1993).  All these \ion{H}{1} observations, separately giving
the edge-on and the face-on views, support the picture of a substantial
disk-halo gas circulation in spiral galaxies (Sancisi, Kamphuis, \&
Swaters 1996) in accordance with galactic fountain models (Bregman
1980). 

Perhaps the strongest evidence linking the substantial disk-halo
\ion{H}{1} flow with the star formation activity in the disk of NGC 891,
comes from the pronounced north-south asymmetry already seen in the
radio continuum and in the H$\alpha$ maps mentioned above and now also
found in \ion{H}{1}.  The considerably higher activity in the northern
part of the galaxy is clearly manifested by the brighter and thicker
radio continuum and H$\alpha$ there.  A similar north-south asymmetry is
now also indicated by the two maps (Fig.\ 3) which show the vertical
position-velocity structure of \ion{H}{1} for the north and the south
sides of NGC 891.  In both maps, the emission seen close to the systemic
velocity and stretching up above 1 arcmin from the plane is from the
slow-rotating halo identified above.  At velocities closer to the
maximum circular velocity, the \ion{H}{1} layer looks clearly thicker in
the north than in the south.  A straightforward interpretation is that
this is due to \ion{H}{1} at intermediate $z$-heights (30-60 arcsec)
which rotates almost as fast as the gas in the plane and is present only
on the northern side of NGC 891 where the star formation activity is
higher.  It could be fresh gas flowing up from the plane and still
connected with the fast rotating, low-$z$ \ion{H}{1}. 

In summary, the similarity between the morphologies of the extended
H$\alpha$, the radio continuum and the \ion{H}{1} is the best
evidence, together with the kinematical information discussed above,
that the \ion{H}{1} is part of the halo and not of an outer flare.

It is not clear what may cause the slower rotation of the halo gas.  The
gravitational field, the structure and energy balance of the
interstellar gas and the magnetic field are all likely to play a role. 
If the velocity structure of the gas at high $z$-distances from the
plane is coupled to the gravitational potential, as in the plane, the
observed decrease in azimuthal velocity with $z$ and the non-corotation
of disk and halo gas is revealing the vertical distribution of matter. 
It can be used, perhaps, to discriminate between disk and spheroidal
mass models.  In this picture the largest decrease in azimuthal velocity
would be expected at small radii, as indeed Fig.\ 5 seems to indicate. 
The disk-halo gas circulation may be understood in the light of galactic
fountain dynamics (Bregman 1980; see also review by Spitzer 1990).  The
gas moves upward from the disk into the halo and radially outward
conserving angular momentum and losing azimuthal speed.  Alternatively,
a large asymmetric drift would also cause a decrease of the rotational
velocity of the halo gas.  Here, however, it seems unlikely to have an
important effect as the required large velocity dispersion ($>$50 km/s)
is not seen in the \ion{H}{1} data, but more sensitive observations
would be needed to rule it out completely. 

As to the flare, it should be understood that the present results do not
rule out the possibility that the \ion{H}{1} layer of NGC 891
may become thicker in its outer parts.  The consequence of this work is
that now the picture has become less simple and that the analysis
required to extract the vertical density profile of the \ion{H}{1} layer
will be dependent on the kinematical model adopted. 

\section{The nuclear disk or ring}

In the central region of NGC~891 there appears to be a depression in the
\ion{H}{1} density distribution (Fig.\ 1), which in itself would not be
unusual in a spiral galaxy of such early type.  But in this case the
effect is enhanced by the absorption against the radio continuum and by
the \ion{H}{1} self-absorption.  This effect is strongest near the minor
axis because of velocity crowding.  Also, note the absorption feature
visible in the position-velocity map (Fig.\ 2 top) between 500 and 600
km/s at the position of the supernova SN 1986J (Rupen et al.\ 1987),
about 1 arcmin to the SW of the center. 

Near the center of NGC 891 the position-velocity map (Fig.\ 2) shows the
presence of faint emission at very low (NE side) and very high (SW side)
radial velocities.  This \ion{H}{1}, not seen in the earlier, less
sensitive observations, follows the pattern of the overall rotation of
the galaxy with approximately the same amplitude.  It has the
unmistakable signature of a fast rotating disk or ring of gas with a
radius of approximately 15$''$ (700 pc) and coplanar with the galaxy
disk.  In the $z$-direction it is unresolved by the $12''$ beam: its
thickness must then be much less than 500 pc.  There is little doubt
that this is the same rotating nuclear disk discovered in CO.  Its
estimated \ion{H}{1} mass is $1\cdot 10^{7}\hbox{~M}_\odot$, with a
factor of two uncertainty.  The radial velocity changes from 300 to 760
km/s.  The maximum rotation velocity is 230 km/s, about equal to that of
the main disk. 

The value of the M(H$_2$) to M(\ion{H}{1}) ratio for the nuclear disk,
obtained using the H$_2$ mass estimated by GGCD, is $\sim 20$, similar
to the value of $\sim 30$ for the Galactic nuclear disk (Combes 1991),
but much larger than that of $\sim 1$ found for the whole disk of NGC
891. 

\acknowledgements

We thank R.  Bottema, F.H.  Briggs, K.H.  Kuijken and T.S.  van Albada
for valuable comments.  The WSRT is operated by the Netherlands
Foundation for Research in Astronomy with the financial support from the
Netherlands Organisation for Scientific Research (NWO).

\newpage

\figcaption[]{ \ion{H}{1} map of NGC 891 superposed on the optical image
(courtesy of Balcells \& Swaters).  The contours are 0.07
($\simeq1.5\sigma$), 0.17, 0.46, 1.1, 2.3, 4.1, 6.4, $9.2\cdot 10^{21}$
\ion{H}{1} atoms cm$^{-2}$ The cross indicates the position of the
central radio continuum source.  The HP beamwidth is $20''$.  The dotted
line indicated the edge of the CCD frame.}
        
\figcaption[]{ Top: Position-velocity map along the major axis of NGC
891.  The dots show the rotation curve (see Fig.\ 5).  The contours are
-0.6, 0.6 ($2\sigma$), 1.2, 2, 5, 10, 15, 20, 25 mJy/beam.  The angular
and velocity resolutions are $15''$ and 33 km/s (FWHM). Bottom:
\ion{H}{1} map of NGC 891 at full resolution ($17.8''\times 11.2$). 
Contours are 0.14 ($\simeq1.5\sigma$), 0.35, 0.9, 2.3, 4.5, 8.3,
$12.9\cdot 10^{21}$ \ion{H}{1} atoms cm$^{-2}$.}
        
\figcaption[]{ Position-velocity maps perpendicular to the major axis
for the north (top) and south (bottom) sides of NGC 891.  They have been
obtained by integrating over strips parallel to the major axis from
$1.25'$ to $2'$ from the center.  The angular and velocity resolutions
are $12''$ and 33 km/s (FWHM).  Contour levels are -5.6, -2.8, 2.8
($2\sigma$), 5.6, 8.4, 15, 30, 50, 80, 110, 140, 170 mJy/beam. 
Velocities are heliocentric.  The vertical dashed line indicates the
systemic velocity of NGC 891 (528 km/s).}

\figcaption[]{ Representative channel maps from different models and
from the observations of the northern (approaching) side of NGC 891. 
The channel radial velocities run in steps of 33 km/s from systemic (528
km/s, heliocentric) at the top to near rotational (299 km/s, hel.) at
the bottom.  The contour levels are -0.62, 0.62 ($2\sigma$), 1.24, 2.5,
5, 10, 15, 20, 25 mJy/beam.  The HP beamwidth is $11.2''\times
17.8''$ The centre of NGC 891 is marked by the cross on the right.}

\figcaption[]{ The rotation curve of NGC 891 for the gas in the plane of
the galaxy (filled circles) and for a strip from 30 to 60 arcsec above and
below the plane (open circles).  Both curves are north-south averages.}

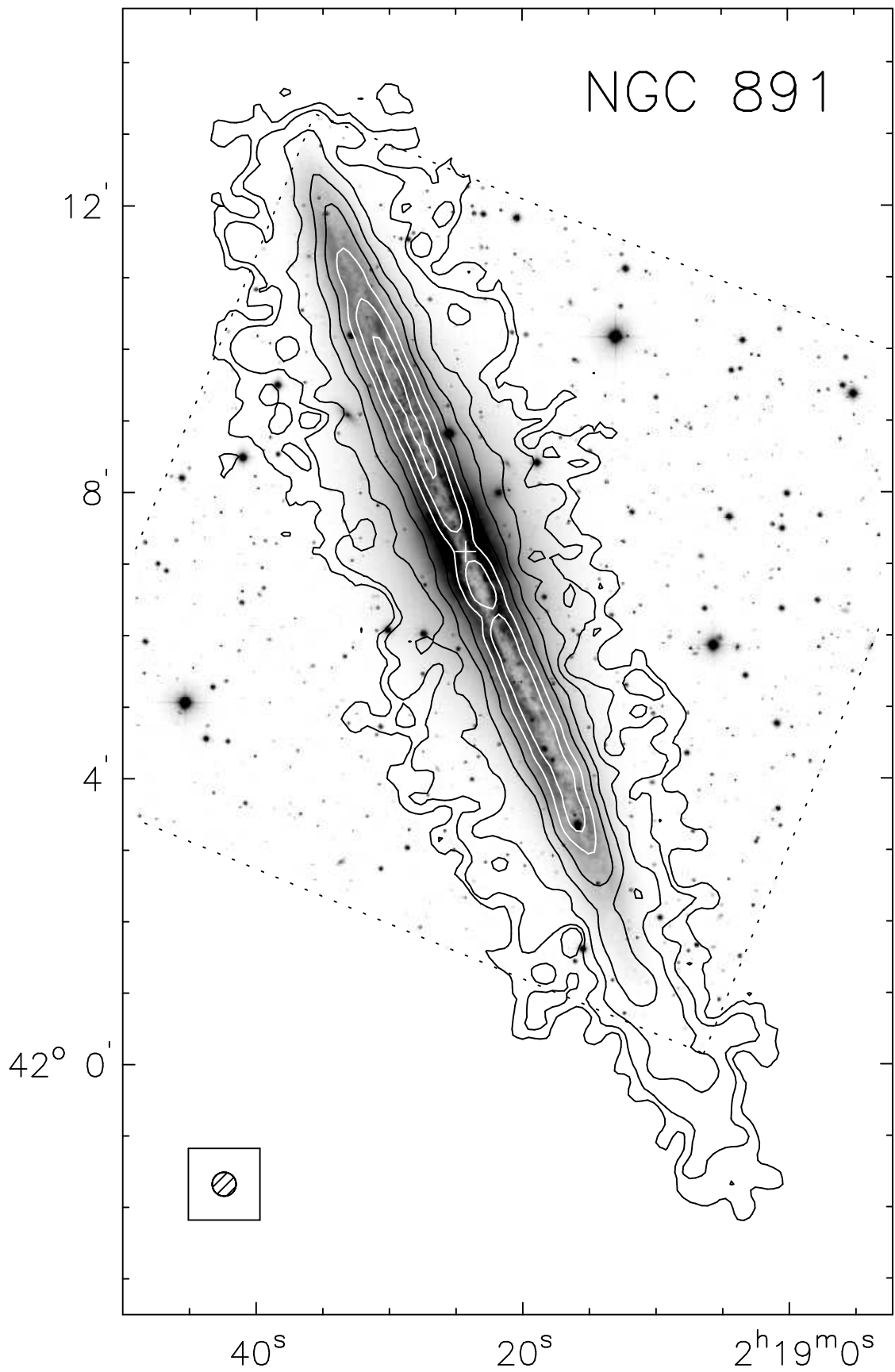
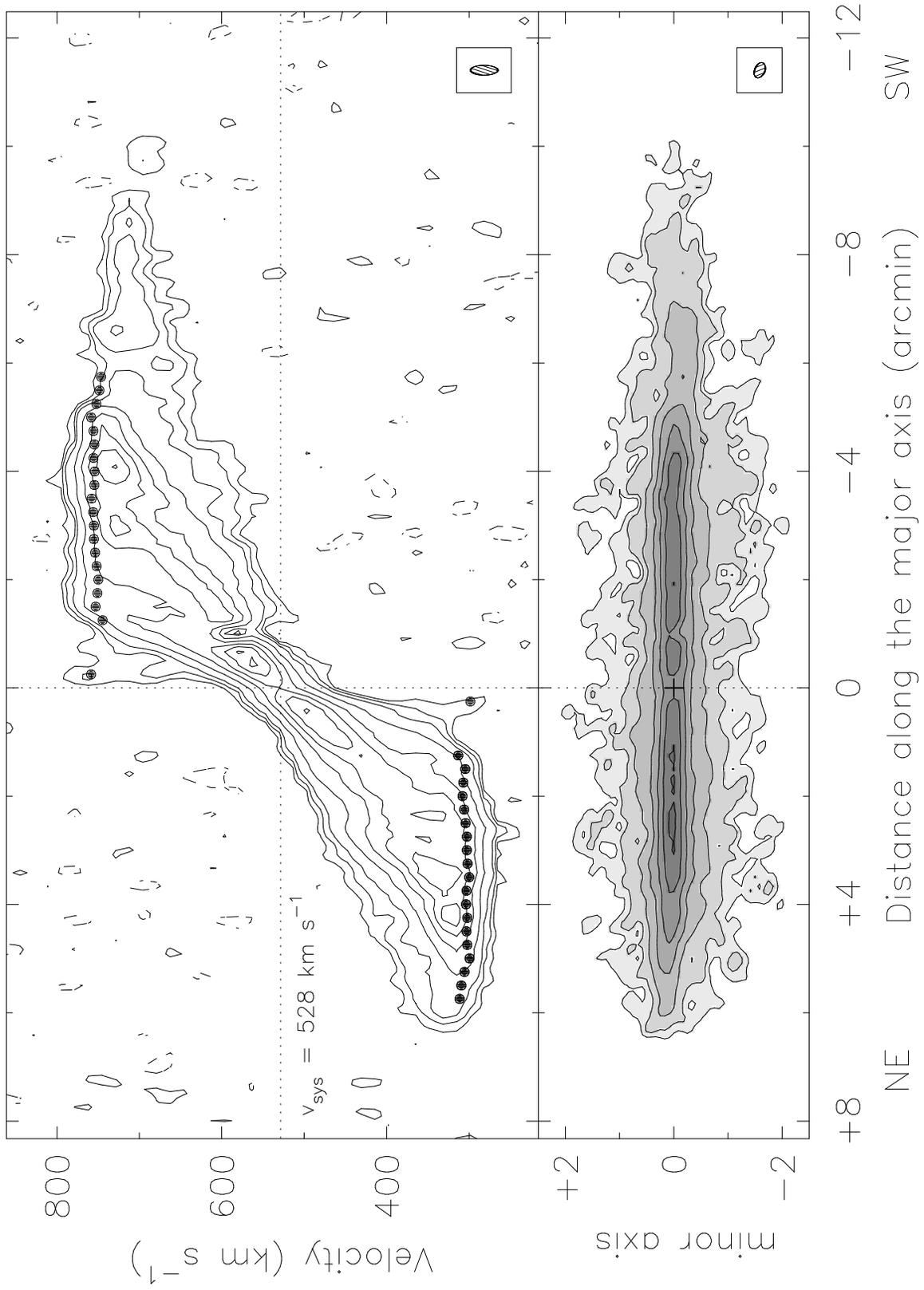
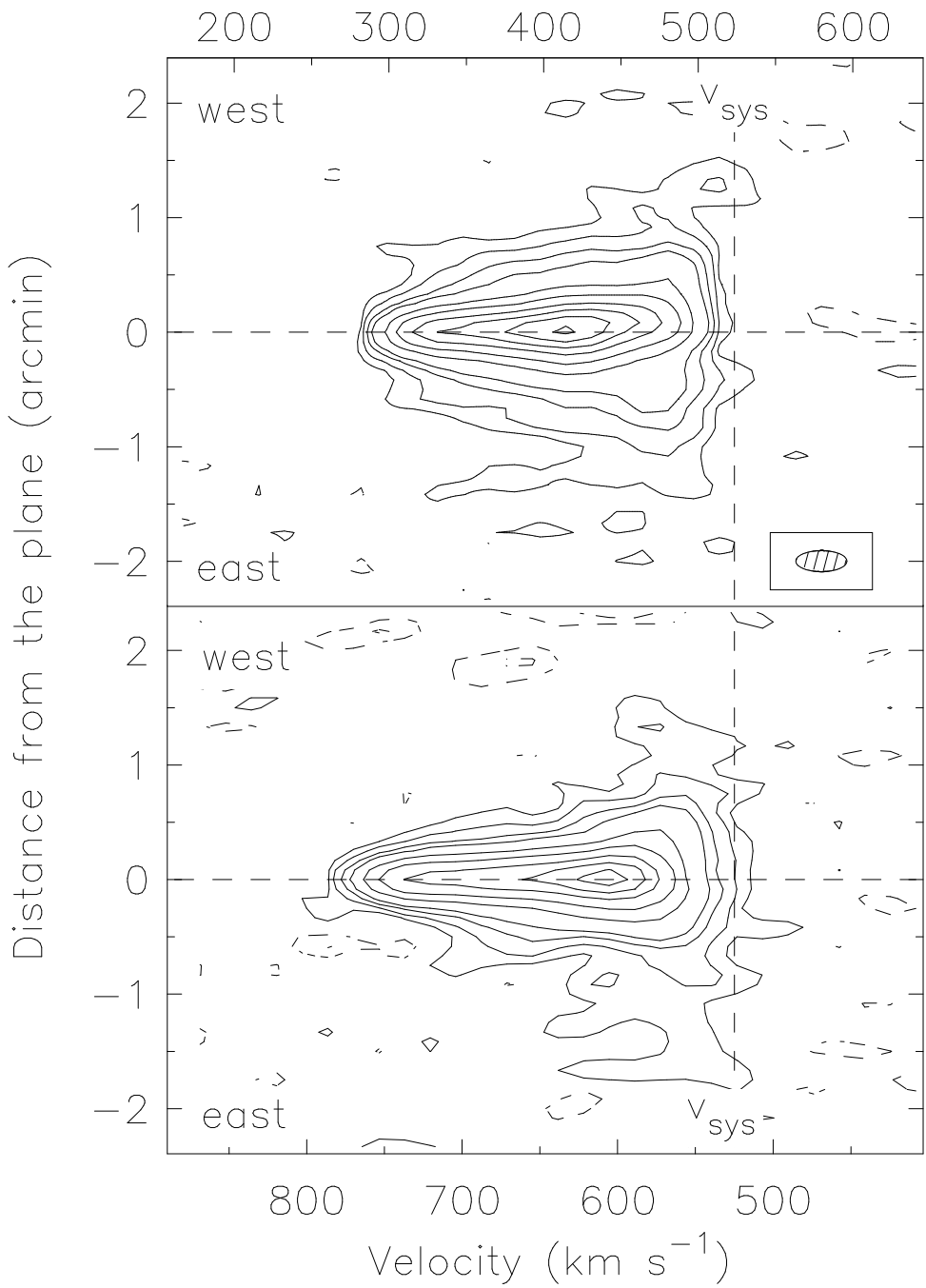
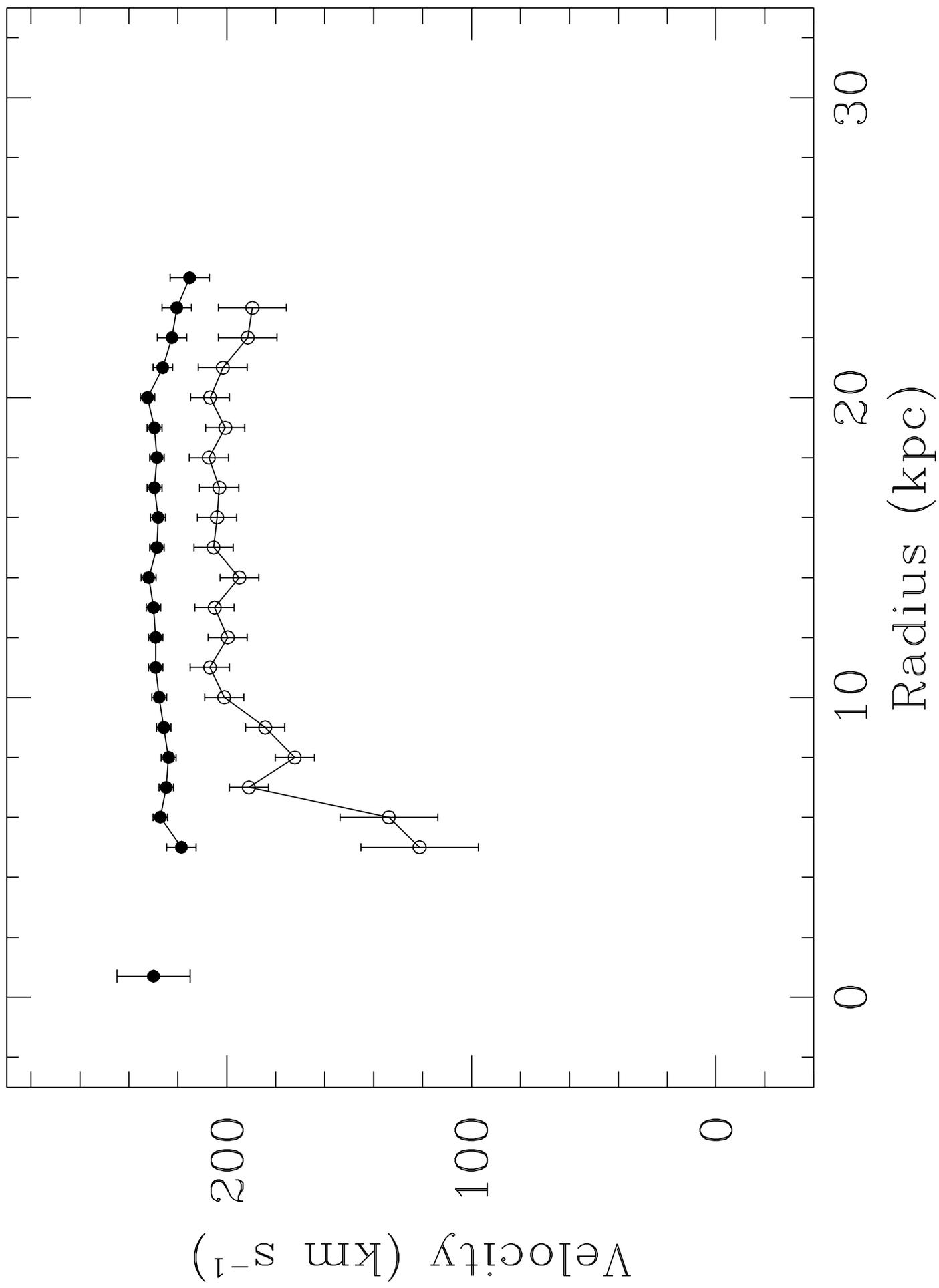
\end{document}